# Words of Wisdom: Representational Harms in Learning From AI Communication

Amanda Buddemeyer[1,2], Erin Walker[1,2] and Malihe Alikhani[2]

[1]*Learning Research & Development Center, University of Pittsburgh, Pittsburgh, PA, USA, 15260*
[2]*School of Computing and Information, University of Pittsburgh, Pittsburgh, PA, USA, 15260*

**Abstract**
Many educational technologies use artificial intelligence (AI) that presents generated or produced language to the learner. We contend that all language, including all AI communication, encodes information about the identity of the human or humans who contributed to crafting the language. With AI communication, however, the user may index identity information that does not match the source. This can lead to representational harms if language associated with one cultural group is presented as "standard" or "neutral", if the language advantages one group over another, or if the language reinforces negative stereotypes. In this work, we discuss a case study using a Visual Question Generation (VQG) task involving gathering crowdsourced data from targeted demographic groups. Generated questions will be presented to human evaluators to understand how they index the identity behind the language, whether and how they perceive any representational harms, and how they would ideally address any such harms caused by AI communication. We reflect on the educational applications of this work as well as the implications for equality, diversity, and inclusion (EDI).

## 1. Introduction

AI communication is widely used in educational tools, including pedagogical robots [1], applications that help to build reading skills [2], or with shorter and simpler language such as hints delivered by intelligent tutoring systems [3]. The language of an AI may be either *generated*, meaning that it is crafted by a statistical model, or *produced*, meaning that it is crafted by one or more human beings. In either case, the language can encode information about culture and demographic identity that users might attribute to the AI. This could be an intentional choice to assign the AI a particular gender, age, or other identity. It could also be unintentional, reflecting identity information about the human(s) who contributed to generating or producing the language. We hypothesize that AI communication has cultural identities whether its creators intend it to or not, which could create harms of representation if the creators give insufficient consideration to those identities and how users perceive them. Harms of representation occur when "a system...represents some social groups in a less favorable light than others, demeans

CEUR Workshop Proceedings (CEUR-WS.org)

them, or fails to recognize their existence altogether" [4, 5]. We study this problem in the context of VQG.

There is reason to believe that all natural language indexes identity information about the person or people who produce it [6, 7]. This extends to include natural language generation models, which are trained on a corpus of produced language from a variety of people. Language that many researchers may assume is "neutral" with respect to ethnic, racial, gender, age, and other identities may simply reflect the identity of dominant cultural groups or otherwise normalize subtle cultural stereotypes. This topic is relevant in many educational contexts, which we explore in Section 2. We examine how a lack of consideration for the identity encoded in language might harm young learners, particularly those from underserved communities.

VQG can be an important application in several educational tasks such as dialogic reading of storybooks and multimodal interaction with situated agents such as pedagogical robots. In this work, we build VQG data sets with mixes of participants from varied racial, gender, and age groups. We present generated questions to a different set of participants from underrepresented or underserved racial, gender, and age groups to interview them about their perceptions of the language. We also explore the applications of this work in technology-enhanced learning by looking at how it might impact the Parent-EMBRACE project [8]. We seek to understand the following questions: 1) How human evaluators index AI communication in terms of identity, 2) How that indexed identity corresponds to the identity of the human being(s) who contributed to crafting the language, either by producing it directly or by contributing to a corpus used to train a generative model, 3) Whether and how human evaluators from underserved or underrepresented groups perceive representational harms from AI communication, and 4) How human evaluators from underserved or underrepresented groups recommend mitigating perceived harms.

After presenting the current status of this case study, we evaluate the implications of this work for EDI, reflecting on how this work promotes EDI as well as identifying aspects of the work that leave unanswered questions or that could be improved.

## 2. Representational Harms in AI Communication

In recent years, many of the technologies and strategies used for AI tasks have been re-examined with an eye toward EDI. The goal of many of these EDI tasks has been to mitigate *harms of allocation* [5, 4] - the type of harm caused when the technology discriminates in how it assigns an outcome to a particular group of users. Minimizing or eliminating harms of allocation can be an important goal of technologies that categorize, retrieve, or predict information such as resumé screening models [9] or prediction models for policing or health care [10, 11]. For communication with AI, however, we must additionally consider *harms of representation* [5, 4], which may occur if the language positions one group as being "normal" while another is "abnormal", if the language positions one group as being superior to another, or if the language reinforces negative stereotypes about a particular group.

When a human being produces language, this is an *act of identity* coded to reflect membership in groups based on race, ethnicity, gender, age, and social class, among other identifiers [6, 7]. Indeed, previous research has found that linguistic behavior that is typically viewed as "unracialized" or "standard" by members of the dominant ethnic or racial group in a society is widely indexed by minorities as being associated with the dominant group [12, 6, 13]. When human speakers or writers express themselves linguistically, they provide clues about their identity; the identity information encoded in the message matches the identity of the speaker or writer. When generated or produced language is attributed to an artificial source, the user may attribute the identity information encoded in the language to that AI rather than to the human producer(s). Without careful consideration, users interacting with AI may index an identity that is inconsistent, that reinforces negative stereotypes, or that normalizes the language of a dominant social group as "standard" or even superior to other linguistic varieties [14, 15].

There is no shortage of educational applications that involve communication with AI, creating potential for representational harms. Stereotypes can negatively impact the cognitive performance of children at least as much as adults [16] and marginalized populations such as girls and learners of color can be particularly susceptible to stereotype threat [17, 18]. We also have reason to believe that the "hegemonic ideologies" perpetuated by holding one linguistic variation in the classroom as being more "standard" or "correct" than others can negatively impact academic performance among marginalized learners [19]. It stands to reason that linguistic behavior that subtly advantages one group over another, creating representational harms, could possibly disadvantage young learners in terms of academic performance, engagement, motivation, self-efficacy, and any number of other metrics. These representational harms are under-explored in educational literature.

The specific task explored by this work, VQG, is pertinent in the field of education. Asking questions is a vital part of education, both as a way for parents and teachers to encourage learners [20, 21] and as a way for learners themselves to consolidate knowledge and engage with the curriculum [22, 23]. It follows that numerous education technologies involve question-asking, including pedagogical robots and agents [24], intelligent tutoring systems [25], or simply applications that encourage learners to be curious [26]. Questions that are generated from visual information can support dual-coded learning using images and language [27], with potential applications in dialogic reading with picture books [8, 28] or with pedagogical robots that generate questions based on visual input.

We propose that it is important for creators of an AI to be deliberate about how the identity behind its language is indexed by users. In the case of pedagogical robots or agents, it's fairly common for creators to assign some sort of cultural identity such as age or gender. However, in the case of unembodied language sources such as intelligent tutoring systems, AI is often assumed to be "without identity". This is a linguistic equivalent of the "view from nowhere" discussed by Gebru in *The Oxford Handbook of Ethics of AI* [29]: AI creators assume an objectivity that may not hold up under careful scrutiny. Even when some identity is attributed to the AI, other aspects of the identity encoded in the language may remain unconsidered. Furthermore, having assigned a certain background or identity to an AI, technology creators may find that users do not index the identity behind the language as the creators intended.

## 3. Case Study

### 3.1. Study Aims

The contributions of our VQG case study includes a dataset of question-image pairs with age, gender, and ethnicity information about the participants. We target several demographic groups that are either historically underserved by AI technology or tend to be underrepresented in the crowdsourcing platforms that often are used to create such corpora. We are unaware of any such dataset, since most do not include demographic information. We have also developed a protocol for interviewing human evaluators that is described in Section 3.4. All human evaluators will be from the same demographic groups targeted for the corpus, which we discuss in more detail in Section 3.2.

The goal of the work is to better understand how human evaluators, particularly those from underserved or underrepresented groups, index the identity behind communication from AI, and whether those perceptions change when we vary either 1) the demographics of the crowdworkers whose data is used in the training corpus (hereafter referred to as the *demographic character* of the model, and/or 2) the demographics of the human evaluators.

### 3.2. Participant Demographics

For our data set of questions, we will collect age, gender, and racial/ethnic information about each participant. We used a crowdsourcing platform called Prolific (https://www.prolific.co). We found Prolific to include tools for screening participants that suited our purpose better than Amazon Mechanical Turk (https://www.mturk.com) (AMT), which was used to collect data for most of the VQG data sets available today. With regard to the demographic categories mentioned in this paper, we found Prolific to have similar representation to what we discuss below for AMT and other crowdsourcing platforms. Our goal is not to collect a data set that is balanced in terms of these demographics. Rather, we seek to create data sets that have a somewhat coherent cultural identity so that we can determine if human evaluators index that identity when they read questions generated by the VQG model.

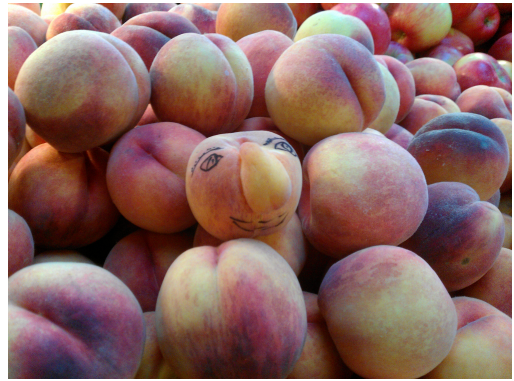

Figure 1: Sample questions for this image might be: 1) Who drew a face on that peach? 2) Is it rude to draw on peaches?

We will gather data from participants in three overlapping demographic groups: participants who self-identify as Black, participants who self-identify as female, and participants who self-identify as being aged 40 or older. We chose these three specific demographics for several reasons.

1. Two of our demographic groups, Black participants and age 40+ participants, are underrepresented in AMT and other crowdsourcing platforms [30, 31, 32], which are over-whelmingly used to train not only VQG models but many other types of generative language models used in the world today. We are interested to see whether and how human evaluators index identities that may not typically contribute to the generated text that they interact with elsewhere in their lives.

2. Two of our demographic groups, Black participants and female participants, are historically marginalized and underserved by many AI applications. We consider it important to explore whether and how human evaluators index these identities.

3. Unfortunately, practical considerations must play a role in our choices as well, and these three demographic groups are all ones for which we could find a sufficiently large group of participants in a reasonable time frame.

Section 4.2 includes further discussion of the tradeoffs of these demographic choices from the perspective of EDI.

### 3.3. Data

We model our data gathering protocol after Mostafazadeh et. al. [33], seeking questions that might be of interest to a human being or spur conversation. Our images come from the MS COCO image data set [34], which was also used by Mostafazadeh et. al. An example of an image with both VQG-style questions can be seen in Figure 1.

In addition to a baseline VQG model trained on a subset of the VQG data set created by Mostafazadeh et. al. [33], we intend to train models on three different data sets, each dominated by data from participants with a demographic trait described in Section 3.2. To maximize resources, we limited our image data set to 500 images randomly selected from among the COCO images used by Mostafazadeh et. al.

We intend to gather data from at least 150 participants in each demographic group, recognizing that some participants are in two or three of the demographic groups. Each participant contributes between 10 and 16 questions, two questions each for five to eight images. Table 1 shows the data that we have gathered as of the writing of this paper. We have gathered data using two surveys in Prolific, one that allows only Black crowdworkers and one that allows only crowdworkers aged 40+. We screen out anyone who does not live in the United States, anyone who does not speak fluent English, and anyone without sufficient vision to be able to view the images. We also screen out crowdworkers who do not pass an attention check. As we gather data, we will pay attention to the number of female crowdworkers who respond to the first two surveys (Black and 40+) and we may add a third survey to gather data from exclusively female crowdworkers.

We will not analyze this data until we have gathered a full data set. Once our full data set is collected, we will analyze it for differences in question length, formality, predicates of personal taste, and other aspects of grammar and usage that may differentiate the language of different groups [6, 35]. Our ultimate interest in this data set, however, is in whether and how human evaluators index identities behind the language.

**Table 1**
Counts of participants from whom we have gathered data thus far. *All participants must self-identify as either 40+ or Black so all participants under age 40 are Black. ** While Prolific allows crowdworkers to specify genders other than female and male, none of the crowdworkers in this data set specified another gender.

|  | Black | | | Race/Ethnicity Other than Black | Unknown |
| --- | --- | --- | --- | --- | --- |
|  | >= 40 | < 40 | Total | >= 40* | >= 40* |
| Female | 13 | 12 | 25 | 25 | 1 |
| Male** | 7 | 14 | 21 | 19 | 1 |
| Unknown | 2 | 5 | 7 | 2 | 0 |
| Total | 22 | 31 | 53 | 46 | 2 |

### 3.4. Interviews

We are inspired by the interviews conducted by Bennett et. al. [36] to understand how human beings are represented in image descriptions. These interviews presented data to interviewees who are most typically harmed by misrepresentation in image descriptions and allowed them to identity problems and discuss potential solutions without being led by the interviewer. Our intention is to follow this model by presenting our human evaluators with examples of images paired with questions generated using models of different demographic character and allow them to reflect on how they index the identity of the anonymous question-asker. After they evaluate a set of images and question pairs, we will provide them with some general information to help them understand how VQG is accomplished using data sets of images and human-produced questions, as well as potential applications for VQG. We will reveal the demographic character behind each image/question pair and ask the evaluators to reflect on potentially helpful and harmful aspects of models of different demographic character. Finally, we will ask them to imagine how VQG might ideally be done to minimize harms.

## 4. Case Study Evaluation

### 4.1. Implications for Equality, Diversity, and Inclusion

This work was conceived as an attempt to promote EDI in technologies that involve AI communication, but it does not offer easy solutions and is far from perfect. In this section, we discuss how the insights of this work might be incorporated into technology-enhanced learning to promote EDI and also how it might create challenges. Our case study partially considers the demographic categories of gender, race, and sex, but language can contain information about other aspects of identity including socioeconomic status, sexual orientation, religious background, and many more. It is not feasible for technologists to consider every possible aspect of identity encoded in language and to ensure that all learners interacting with that technology feel validated and included. Indeed, if the interviews in this work have results similar to the interviews that they are modeled after, published by Bennett et. al. [36], members of the populations marginalized by this issue will likely come up with solutions that directly contradict each other. Communities do not speak with a single voice. However, even just being aware of the possibility of representational harm in AI communication can lead technology creators to be more considered in their approach, especially when creating educational technology.

One potential strategy to address, if not necessarily mitigate, representational harms in AI communication is to teach young learners to recognize how identities can be encoded in language and to foster in them a feeling of empowerment to shape technology to meet their needs and those of their communities. This is a topic addressed by Buddemeyer et. al. [37] in a project intended to create a programmable dialog system paired with a culturally responsive curriculum intended to promote reflection on the connection between language and identity [38]. The curriculum encourages learners to envision how they might shape the dialog system to fit their needs and the needs of their communities, building computational thinking skills and an agentic relationship with technology [38].

## 4.2. Implications for Different Subgroups

It is not feasible to create language that makes every child in every classroom feel validated and included. We spoke in Section 2 about the impossibility of creating language that is "neutral", which naturally leads to a question: what identity should be encoded in AI communication?

The answer may seem obvious for some applications. The Parent-EMBRACE application discussed in the Introduction, for example, is targeted to Latinx children aged 5-10 in the United States who are learning to read in English, as well as the parents of those children. The creators of this system could simply build personas that represent the most common users and seek to match the language of the system with the users. However, even in the case of such a seemingly-specific demographic group, that strategy could leave many users feeling excluded. Latinx Americans have origins in a variety of countries and regions, including Mexico, Puerto Rico, Dominican Republic, and many others, or a combination. Some come from a rural region, others urban. They have many genders and sexual orientations. The point is, even applications with a seemingly-specific target demographic have users with a wide variety of intersectional identities. The language of an AI cannot incorporate all of these identities.

Beyond the problem of how to select an identity for an AI, there is a further problem of how to implement the selected identity. For applications that use language produced by a human being, there is a delicate balance in invoking the identity authentically, avoiding what Buchotz and Lopez refer to as *linguistic minstrelsy* [39]. For applications that use generated language, technologists must be able to build a sufficiently large data set to train a model. This can be an expensive and time-consuming process if one is gathering data outside a crowd-sourcing platform like AMT or Prolific. When using a crowd-sourcing platform, there may be limited data on demographic groups that are small, underserved, or underrepresented. For example, it would have been valuable to this work to create a corpus of VQG questions from participants with non-binary gender, but fewer than 2,000 Prolific workers had a gender other than male or female, and that was before applying our additional exclusion criteria, such as requiring crowdworkers who lived in the United States, spoke fluent English, and had sufficient vision to view images. We simply were not confident that we could gather enough data from that group in a reasonable time frame. Similar statements could be made about ethnic groups more specific than "Black" or more intersectional identities such as "Black females".

With crowdsourcing platforms, one must also contend with selection bias. For example,

the gender options in Prolific are: *Male, Female, Trans Male/Trans Man, Trans Female/Trans Woman, Genderqueer/Gender Non Conforming, Different Identity, Rather Not Say*. It's clear that the designers of Prolific are trying to be inclusive, but this list forces potential participants who are transgender to select an option that excludes them from their own gender. For instance, selecting "Trans Male/Trans Man" here implies that the participant is not male. Another obvious problem comes with the list of ethnicities, which only allows the user to select one ethnic identity. Potential participants may decide not to participate because they feel that their identity is being misrepresented or disrespected. The point here is not to attack Prolific or AMT or any particular crowdsourcing platform but rather to point out that inclusion is an ongoing process that requires feedback and iteration to always do better. We chose to gather our data on a crowdsourcing platform, which was a much faster and more convenient way to recruit participants to build our question corpus than using more traditional methods. The benefits of crowdsourcing are not small: many researchers have limited funding, limited time, and must produce results to make progress in the field. It's possible that less EDI research would occur without crowdsourcing. However, we must be aware of the trade-offs that we are making when we choose to use this recruitment method.

## 5. Conclusions, Open Questions, and Future Work

Communication attributed to an AI rather than a human source can encode identity information about the human beings who created the language. This is true of produced language, meaning language created directly by a human or group of humans, and generated language, meaning language created by a natural language generation model based on a training corpus of language created by humans. This identity information can create harms of representation if it positions one group as "standard" or "normal" while another is "other", if it positions one group as superior to another, or if it reinforces negative stereotypes. In this work, we reflect on the implications of this potential problem on young learners, particularly those from marginalized populations, who may be particularly susceptible to stereotype threat and who may suffer academically when they are positioned as "other". We present a case study that examines potential representational harms in VQG. We introduce a new VQG data set that includes demographic information, which can be used for training machine learning models. Our contributions also include a protocol for interviewing human evaluators to better understand how identities are indexed in the generated questions, whether and how evaluators from underserved or underrepresented communities perceive representational harms, and how the evaluators would ideally address representational harms in the future.

The main open question raised by this work is how to move forward given the potential messiness of creating AI communication. AI communication that seeks to be "neutral" can simply position dominant groups as "standard", furthering representational harms. This strategy can also result in language that has inconsistent identity, which can be confusing for users. AI communication that is identifiable as belonging to one group rather than another risks excluding users who are not represented, reinforcing negative stereotypes about the identities represented by the language, and/or promoting linguistic minstrelsy. Regardless of the strategy chosen by technology creators, creating language that indexes a particular identity can be time consuming and expensive.

In an effort to minimize representatinal harms caused by AI communication, we encourage researchers and educators to consult with impacted communities about potential mitigations, iterate on different strategies, and build curricula around problems that are difficult to solve so that learners become part of each successive step of improvement.

## Acknowledgments

Thanks to the anonymous reviewers from LearnTec4EDI.